\documentclass[sigconf]{acmart}
\AtBeginDocument{%
  }

\setcopyright{acmlicensed}

\copyrightyear{2027}
\acmYear{2027}
\setcopyright{cc}
\setcctype{by}
\acmConference[GROUP Companion '27]{The 2027 ACM International Conference on Supporting Group Work}{January 10--13, 2027}{St. Simons Island, GA, USA}
\acmBooktitle{The 2027 ACM International Conference on Supporting Group Work (GROUP Companion '27), January 10--13, 2027, St. Simons Island, GA, USA}
\acmDOI{10.1145/3838685.3849860}
\acmISBN{979-8-4007-2916-4/2027/01}

\begin{document}

\title{Measuring “Success” In the Transition from Pre- to Post-LLM Conversational Agents through the Lens of Interaction \& Explanations for Older Adults}

\author{Tamara Zubatiy}
\email{t.zubatiy@northeastern.edu}
\affiliation{%
  \institution{Northeastern University}
  \city{Boston}
  \state{Massachusetts}
  \country{USA}
}

\author{Niharika Mathur}
\authornote{This work was conducted while the author was affiliated with Georgia Institute of Technology.}
\email{nmathur9@jh.edu}

\affiliation{%
  \institution{Johns Hopkins University}
  \city{Baltimore}
  \state{Maryland}
  \country{USA}
}

\author{Reina Chan}
\email{chan.szey@northeastern.edu}
\affiliation{%
  \institution{Northeastern University}
  \city{Boston}
  \state{Massachusetts}
  \country{USA}
}

\author{Srijan Pokharel}
\email{pokharel.sr@northeastern.edu}
\affiliation{%
  \institution{Northeastern University}
  \city{Boston}
  \state{Massachusetts}
  \country{USA}
}

\author{Rosemarie Santa González}
\email{rgonzalez308@gatech.edu}
\affiliation{%
  \institution{Georgia Institute of Technology}
  \city{Atlanta}
  \state{Georgia}
  \country{USA}
}

\author{Agata Rozga}
\email{agata@gatech.edu}
\affiliation{%
  \institution{Georgia Institute of Technology}
  \city{Atlanta}
  \state{Georgia}
  \country{USA}
}

\author{Xiang Zhi Tan}
\email{zhi.tan@northeastern.edu}
\affiliation{%
  \institution{Northeastern University}
  \city{Boston}
  \state{Massachusetts}
  \country{USA}
}

\author{Elizabeth D. Mynatt}
\email{e.mynatt@northeastern.edu}
\affiliation{%
  \institution{Northeastern University}
  \city{Boston}
  \state{Massachusetts}
  \country{USA}
}

\renewcommand{\shortauthors}{Zubatiy et al.}
\renewcommand{\shorttitle}{Measuring "Success" in Conversational Agents}

\begin{abstract}
  Conversational agents (CAs) for older adults have moved through at least two distinct eras, split by the emergence of large language models (LLMs) at the end of 2022. The first era was defined by commercial smart speakers such as the original Google Home and Amazon Alexa, built on rigid, rule-based dialogue. Interactions were single-turn, user-initiated and "successful" if and only if a command matched one of a finite set of pre-templated intents. The current era is marked by LLM-powered systems capable of open-ended, multi-turn conversation, with capacity for personalization and foundations for more proactive interaction. This shift has not only transformed what CAs can do for users, but also changed what researchers can measure when it comes to understanding “success” of user interactions with CAs. In this case study, we reflect on over 8 years of research spanning more than 100 participants across multiple field sites to trace how our own working definitions of “success” have evolved. We reflect on two completed bodies of work with pre-LLM CAs and an in-progress study with a post-LLM CA. We discuss how our definition and measurement of success have evolved during this transition and also ask open questions to the community about how to optimally measure “success” in the post-LLM CA era. This research is deeply applied and subject to the messiness of real-world deployments. Our intended contributions are deliberately practical, a combination of methodological considerations and reflection questions to support researchers engaging with this new wave of technologies.  
\end{abstract}



\maketitle

\section{Introduction}

In this case study, we discuss our experiences of conducting research through the transition from pre- to post-LLM Conversational Agents (CAs) and the impact of this transition on user experience research, specifically, on the metrics used to quantify success. We explore this transition through reflecting on 2 case studies across pre- and post-LLM CAs. For each case, we explain how we quantified and measured our participants’ success, and also share the various limitations we encountered. These limitations are the result of the interconnectedness of the inherent limitations of the technologies and an inability to find a quantitative metric that accurately characterizes success for Conversational Assistants. To showcase the methods \& limitations of the pre-LLM era, we first present \textbf{Cases 1a and 1b}, introduced in the next subsection of this introduction. The rapid rise of LLMs in commercial CAs introduced new promise for research but also posed challenges, from privacy concerns to hallucination in responses \cite{zhang_its_2024}. To explore this new era, we are currently embarking on two new studies, described in \textbf{Case 2}, which build on our prior work with pre-LLM systems, but in this new LLM era. In this paper, we explore the methodological changes we have implemented in response to the post-LLM era and propose ideas for researchers navigating similar research questions.


\section{Case Study 1: Pre-LLM Deployments}

\subsection{Introduction to Study Setup (Case 1a and 1b)}

For Case 1a, we specifically consider the work from two of our previous papers -- (1) "Empowering Dyads of Older Adults With Mild Cognitive Impairment and Their Care Partners Using Conversational Agents" (n=10 dyads, 10-week single deployment) \cite{zubatiy_empowering_2021}; and (2) "'I don't know how to help with that' — Learning from Limitations of Modern Conversational Agent Systems in Caregiving Networks" (n=26 dyads across three successive 10-week deployments, which include and extend the sample in \cite{zubatiy_empowering_2021} as its first cohort) \cite{zubatiy_i_2023}. The combined studies considered 9328 interactions and 26 summative, post-deployment interviews. For the studies in the pre-LLM environment, we defined interaction as a user-initiated command to the system OR a user-initiated command that was triggered at a time-delay (i.e., setting up an alarm/reminder). 

These studies share a common deployment context and methodology across the same population of older adults with MCI and their care partners. Each deployment involved the following protocol: we set up a commercial CA, deployed for 10 weeks in the homes of dyads composed of an older adult with mild cognitive impairment (pwMCI) and a care partner (CP). Similar to a technology probe \cite{hutchinson_technology_2003}, there was no explicit, researcher agenda. The goal was to capture older adults' naturalistic interactions with the CA and measure both the \textbf{number and quality of interactions}. The data analysis methodology was quantitative log analysis and qualitative semi-structured interview coding. For categorizing the interactions logged, we leveraged a framework inspired by Bentley et al. in \cite{bentley_understanding_2018}. They categorize all possible user commands to the system as one of three categories: Level 1 "out-of-the-box”, such as asking about the weather; Level 2 "personalized", such as setting a reminder or calendar appointment; and  Level 3 "skills", third-party apps like games. Both papers deliberately focused on the pwMCI-CP pair, or dyad, as a single unit, which allows us to consider the benefits to and dynamics within the care system,  rather than just one individual. The CA system was conceptualized and positioned as a compensatory technology as illustrated by the “filling cup” metaphor in \cite{zubatiy_i_2023}, suggesting that the CA is there to fill a widening gap in care as the pwMCI loses their functional capabilities, and the CP (often an older adult themselves) is increasingly unable to keep up. This gave us an implicit definition of success as keeping up with the needs of the person with MCI over time through functional support.

For Case 1b, we introduce a sub-deployment, conducted as a constituent part of the two larger deployments in Case 1a, and consider the work in our previous paper -- "A Collaborative Approach to Support Medication Management in Older Adults with Mild Cognitive Impairment using Conversational Assistants (CAs)" (n=7 dyads, 20-weeks) \cite{mathur_collaborative_2022}. As a continuation of the deployments in 1a, we conducted this sub-deployment to explore the role of CA-driven support for the specific task of medication adherence, through a conversational system developed within the CA. This reminder functionality was deployed with a subset of users from the deployments in 1a, and leveraged the same home-based deployment context, involving older adults with mild cognitive impairment (pwMCI) and their care partners. Unlike the earlier deployments, which broadly examined how CAs were used in everyday life, this sub-deployment focused on a specific, high-value task, i.e., supporting medication adherence. Throughout the day, the system proactively provided medication reminders, periodically checked with users about their medication status, and encouraged timely adherence through conversational interactions. This deployment provided an opportunity to examine how conversational reminders from CAs could shape users' understanding, trust, and engagement with the system.

\subsection{Case 1a: How Did We Measure Success?}

In the pre-LLM Conversational AI era, the dominant quantitative logic, borrowed directly from the prior smart-speaker usage benchmark literature \cite{bentley_understanding_2018, porcheron_voice_2018}, was that a successful deployment of a CA with real users would be marked by high rates of interaction with the technology. The headline metrics that we looked at were raw interaction counts with the CA and daily/weekly interaction rates per person. Success in each subsequent deployment study was measured by comparing interaction counts to each previous one. Our work published in \cite{zubatiy_i_2023}  similarly reports success metrics in terms of the average number of daily interactions per user. 

In addition to the number of interactions, we measured the complexity of each interaction. We categorized all logged interactions according to the \textit{Levels} taxonomy first presented in the CHI 2021 paper \cite{zubatiy_empowering_2021}. We initially conceptualized this as “non-functional” usage, we found that both CPs and pwMCI most utilized “non-functional” level 1 interactions, like using the CA for entertainment. The level 2 interactions that we observed for older adults with MCI were largely enabled by the care partners. In this case, higher interaction count for the CP indicated success for the pwMCI even if the pwMCI never interacted with the system at all. That being said, care partners also described a sense of relief from caregiving burden and offloaded care and supervision when they were able to successfully set up a desired function using the CA \cite{zubatiy_i_2023}. There were other mismatches between the quantitative and qualitative findings. We also observed participants with higher usage sometimes expressed discontent because the high number of interactions was actually an artifact of repeated failed attempts at getting the CA to do the desired task. We expand on this observation further with Case Study 1b in Section 2.4. Our data analysis across all three deployments deliberately triangulated usage data with self-report interview \& survey data, but it was becoming increasingly clear that we were not successfully measuring success and that each stakeholder had a different definition. 

Another notable reflection was that in these studies, we associate anthropomorphization of the CA as qualitative evidence of the CA’s successful social and emotional integration into the routines of daily life. In the CSCW 2023 work \cite{zubatiy_i_2023}, we similarly observed the treatment of and dialogue toward the CA as a being rather than a machine. The language used by participants indicates their desire for the CAs to be more like conversation partners or peers: animated and full of character. Furthermore, the failure of pre-LLM CAs to respond appropriately to user requests is a leading marker of failure in the post-LLM CA world, both because it is a sign of abandoning the user and not providing long-term persistent memory to drive personalization. At this time, it was simply “unsuccessful” in a binary sense, but there was no concept of what it could be like to interact with CAs outside of the rules of the rigid templated command-style interactions. 

\subsection{Case 1a: Key Limitations of Pre-LLM CAs}

We identified many limitations of pre-LLM CAs that relate to the mismatches between what users reported, what they wanted, and what we measured. In \cite{zubatiy_i_2023}, we identified several of these limitations in the form of tradeoffs: the tradeoff between \textit{flexibility vs. efficiency}, the tradeoff of \textit{proactivity vs. autonomy}, and the tradeoff of \textit{personalization vs. privacy}. In addition to these tradeoffs, we identified that it was quite hard to learn how to use the system and that care partners tended to absorb labor for their partners with MCI rather than attempting to keep teaching the pwMCI to remember the rigid templated commands. 

The flexibility vs. efficiency tradeoff speaks to the tension between wanting to be fluid in your speech to the CA but potentially being limited by the CA’s inability to understand anything beyond a handful of templated command-style user intents, even if efficiently. We also found that this templated style of interaction was susceptible to errors created by older adults' speaking style, which tended to be more conversational and fluid. From a measurement perspective, each error was technically counted as an interaction, creating a misleading signal of “success’ that was actually repeated failure in reality.

The proactivity vs. autonomy tradeoff is particularly significant because, in many ways, it was a constraint of our experimental design. We chose to use the native Google Home functionality for these deployments rather than any custom apps or skills. As a result, we were constrained to the status quo mental model for commercial CAs, i.e., the user should initiate all interactions. This lack of proactivity, in turn, made it more difficult for older adults to gain value out of the system because of factors already discussed above. This created a frustrating loop where users had infinite autonomy but often could not benefit from the system. 

When reflecting on the tradeoff of privacy vs. personalization, we were once again in tension between user-defined and researcher-defined success. Whereas users wanted to have the systems respond empathetically and be aware of previous conversation history, researchers were wary of the privacy implications of sharing personal data with Google. We instructed users to not reveal key personal details to avoid the risk of them going to Google. In reality, there was no persistent system memory of the user for the CA. The CA did not reference information learned in previous interactions and was essentially a blank slate for each conversation, lacking empathy for or knowledge of the user. This resulted in many failures to meet user expectations, including once when a user shared that it was their 49th wedding anniversary and asked for gifting suggestions, and the CA replied “I don’t know how to help with that” \cite{zubatiy_i_2023}.

\subsection{Case 1b: How Did We Measure Success?}

While the deployments in Case 1a highlighted the limitations of measuring conversational success through interaction counts and task completion alone, the sub-deployment in Case 1b involving medication reminders also revealed another related challenge. In analyzing interaction logs, we observed that users were often left uncertain about \textbf{why} the system generated a particular response or reminder at a specific time. Pre-LLM CAs could complete a task that was invoked or set up ahead of time, but they were unable to communicate the reasoning behind their responses in ways that aligned with older adults’ routine, expectations and experiences \cite{enam_artificial_2025, gleaton_taxonomy_2026, mathur_sometimes_2026}. For example, in response to the CA providing a reminder for medication, users would ask questions that reflected their desire to know about the CA's internal reasoning that triggered that reminder, such as “Are you sure I have not taken it [..]? I think I have...”. 

Recurrence of such responses reflected an inability of pre-LLM CAs to provide adequate \textbf{explanations} for their responses to users, and prompted us to dig deeper into the data to understand how users were responding when their explanatory needs were left unmet, and whether they provided any follow-up information to the CA after it failed to deliver an explanation. This inquiry also led us to expand our notion of conversational success beyond successful task completion to encapsulate successful explanations and reasoning as well.

Building upon the interaction-focused metrics discussed in the previous deployments, our initial approach to evaluating the reminders in this deployment initially focused on user engagement through conversational interaction. We anticipated that a successful deployment would be reflected through sustained conversational engagement and users' willingness to interact with the CA during these medication reminders. Interaction logs therefore became our primary quantitative data source for understanding how participants responded to the system over time.

However, a closer examination of these interaction logs revealed a more nuanced picture. Similar to our earlier deployments, higher interaction counts did not necessarily indicate successful interactions. Instead, many conversations became prolonged because the CA failed to adequately address users' questions and their need for more explanations from it. Users frequently attempted to repair these conversational breakdowns by repeating their responses, rephrasing their requests or providing additional context that they believed would help the system better understand their situation. In many cases, users explicitly sought explanations by asking "why" questions in an effort to understand the rationale behind the reminder or recommendation. Our analysis showed that approximately 38\% of all user interactions arose from these conversational misalignments rather than successful exchanges \cite{mathur_it_2025, mathur_collaborative_2022}. Rather than representing productive engagement, these additional conversational turns often reflected users' efforts to compensate for the CA's limited ability to explain its reasoning or adapt to evolving circumstances. Participants volunteered contextual information about changes in their medication schedules, their use of pill organizers or other aspects of their medication routines, implicitly treating the CA as though it should incorporate this information into its future reasoning. However, the pre-LLM system was unable to meaningfully leverage this contextual information or respond with explanations grounded in users' lived experiences.

\subsection{Case 1b: Key Limitations of Pre-LLM CAs}

While our quantitative analysis revealed the prevalence of conversational breakdowns, post-deployment interviews provided a richer understanding of why these breakdowns occurred. Participants consistently described situations in which the CA lacked awareness of important contextual information already present in their daily lives, including changes to medication routines, conversations with caregivers, and information available through everyday artifacts such as pillboxes, calendars and wearable devices. These qualitative findings also shifted how we conceptualized successful explanations. Rather than simply providing a rationale for a reminder, participants expected explanations that were grounded in their personal goals, routines and lived experiences. This realization prompted us to move beyond viewing explanations as static system outputs toward a contextual explanation framework that incorporated multiple sources of conversational evidence, including user history, environmental context, activity information and system reasoning \cite{mathur_categorizing_2024}. Looking back, this work revealed that older adults' explanatory needs extended well beyond what pre-LLM CAs were capable of supporting, foreshadowing many of the opportunities that would later become possible with LLM-based conversational systems.

This sub-deployment represented an important methodological turning point in our research. While our earlier deployments had already begun to challenge the assumption that greater interaction frequency necessarily reflected greater success, this sub-deployment provided further evidence that this assumption was fundamentally incomplete. By digging deeper into interaction logs and triangulating them with participants' qualitative experiences, we found that many additional conversational turns were not indicators of engagement, but rather symptoms of unmet explanatory needs, conversational breakdowns and users' attempts to repair interactions by supplying missing context. These findings prompted us to reconsider the metrics by which we evaluated conversational agents, shifting our focus away from interaction quantity alone and toward the quality of users' understanding, the contextual appropriateness of explanations, and the extent to which pre-LLM CAs were able to meaningfully support user goals.

\section{Case Study 2: Post-LLM Deployment}

For Case Study 2, we reflect on our current work to understand how older adults use post-LLM CAs while living independently in apartments within an assisted living community. As this study is still in the early stages of deployment, we use it as an opportunity to reflect on how the emergence of LLMs has fundamentally reshaped our understanding of what constitutes success in CAs and how such success should be measured. 

\subsection{Framing the Study and User Experience}

For this ongoing work, we are deploying Amazon's Alexa Echo Dot devices into apartment homes within an assisted living community for older adults living independently. The Alexa device has built-in LLM capabilities that allow it to now be more conversational and "chatty", rather than exemplify the rigidity of the pre-LLM CA paradigm \cite{al-amin_history_2024}. To support this deployment, we developed a custom Alexa Skill, called \textit{Robin}, which interfaces with an LLM hosted on our HIPAA-compliant backend rather than relying on Amazon's native conversational capabilities. Participants therefore interacted primarily with Robin for study-specific functionality, while retaining access to Alexa's standard features. This architecture allowed us to separately capture and analyze interactions with Robin alongside participants' broader interactions with Alexa.

These design decisions reflect several important shifts in how we now conceptualize CA research. First, the unit of analysis has transitioned from the caregiving dyad to the individual, as participants in this deployment are independently living older adults rather than individuals living with mild cognitive impairment. Consequently, we no longer frame CAs primarily as compensatory technologies that fill gaps in caregiving. Instead, Robin is designed as a proactive conversational partner that augments and enriches users' everyday experiences, moving beyond the "filling the cup" metaphor that guided our earlier work. Building on the lessons from Cases 1a and 1b, Robin also incorporates a substantially richer understanding of users by drawing upon multiple sources of contextual information, including activity schedules from the assisted living community, individualized user profiles capturing hobbies and preferences and a proactive check-in system. These check-ins occur twice daily at participant-selected times and can incorporate personalized content such as weather updates, scheduled community activities, reminders and other user-specified information. These design decisions reflect a broader shift from reactive, command-driven CAs toward proactive, context-aware systems that are capable of delivering more personalized and meaningful interactions.

\subsection{Case 2: How Should We Measure Success?}

In terms of our methodology and quantifying “success”, we were previously counting interactions with the CA and using a taxonomy to categorize user interactions into 3 levels. As we began this new deployment, we naturally adopted the same approach, developing an automated dashboard to track and categorize each participant's interactions with Robin. However, we now have to consider the implications of the new study design: how do we categorize the proactive check-ins? What if a user doesn’t respond to a check-in? Do they have to respond if they got all the information? Is it not success if they didn’t even have to ask? 
These questions emerged almost immediately during deployment. During one onboarding session, a participant asked, \textit{"Do I have to do anything or will it just work?"} This simple question highlighted the tension between user-centered and researcher-centered notions of success. From a user experience perspective, a CA that seamlessly anticipates and supports users' needs with minimal effort represents an ideal interaction. Yet, under traditional evaluation metrics, fewer interactions might instead be interpreted as lower engagement. This realization reinforced the need to rethink our evaluation methodology so that our definitions of success better reflect participants' experiences and desired outcomes, rather than relying primarily on interaction frequency as a proxy for successful conversational engagement.

Another important shift is that CAs can now sustain open-ended conversations, enabling greater personalization and allowing users to interact with the system more like a conversational partner than a command-driven assistant. Whereas conversations in our earlier deployments typically consisted of a single user command, we now observe interactions lasting over five minutes with dozens of conversational turns. This raises new methodological questions: how should these conversations be categorized? Are they more or less valuable than traditional "functional" interactions such as reminders or alarms? More fundamentally, if reminders and scheduling have become largely automated, is conversation itself becoming the primary outcome of interest?

To help us answer some of these questions, we have transitioned our qualitative data collection to weekly check-ins with participants rather than end-of-deployment reflections. We are also looking more at the number of turns in a conversation, rather than simply the number of user-initiated conversations. We are also reflecting on whether it is even necessary to quantify the number of user-initiated vs. machine-initiated conversations, or whether a better metric would be overall satisfaction with the system and other qualitative data points. 

\subsection{The Impact of LLMs on Previous Key Limitations}

The advent of LLM-powered CAs has empowered us to overcome many limitations we identified in our prior work, but LLMs are not a magic wand, and there is still a lot to figure out when it comes to defining success in the context of these limitations. 

The flexibility-efficiency tradeoff has been fundamentally reframed by LLMs. Users are no longer constrained by rigid, templated commands and can communicate naturally with the CA. However, efficiency remains a concern, as sophisticated LLMs are now used for relatively simple tasks such as scheduling reminders, taking up astronomical amounts of water and energy. More importantly, the question has shifted from how users issue commands to whether they need to issue commands at all. Much of the feedback from Case 1a centered on the repetitive labor of creating and maintaining scaffolding through repeated commands. In contrast, Robin raises the opposite question: should the system proactively anticipate users' needs? As a result, interaction frequency alone is no longer an adequate measure of success. A participant may achieve their goal with few or no user-initiated interactions, making qualitative methods essential for understanding the user experience.

The proactivity-autonomy tradeoff has similarly evolved. While most commercial CAs still follow a "don't speak unless spoken to" paradigm, Robin delivers proactive check-ins at participant-selected times, incorporating personalized content such as weather, community activities and reminders. Users can also add new reminders at any time, balancing proactive support with user control. This design introduces new methodological challenges: participants may benefit from a proactive interaction without responding, resulting in zero recorded user interactions, or they may miss the interaction entirely. Consequently, quantitative interaction logs alone cannot distinguish between successful support and missed opportunities, further reinforcing the importance of qualitative inquiry.

Finally, the personalization-privacy tradeoff has likewise shifted. By hosting Robin's LLM within a HIPAA-compliant university environment, participants can safely share personal information, enabling richer, more contextual interactions. Robin further supports personalization through user profiles that incorporate individual preferences and routines into responses. Yet these new capabilities also raise participants' expectations. As an example, one former English teacher in our participant pool remarked that Robin's discussion of her book felt like "a student who used SparkNotes", illustrating that personalization is now judged by the depth and authenticity of conversation, not simply the presence of personal information. These findings suggest that while LLMs enable unprecedented personalization, evaluating its success increasingly depends on understanding participants' lived experiences and expectations rather than interaction metrics alone.

\section{Key Takeaways and Contributions}

Across both case studies, we set out to measure “success” and across both, our metrics revealed a failure to accurately capture success as defined by users. In the pre-LLM era, the three tradeoffs we identified were not independent design tensions as much as three faces of a broader mismatch between the outdated command-style interaction model and older adults’ expectation of a conversational partner. LLMs did not resolve the mismatch as much as reveal still more complexity. Flexibility is not scarce, but efficiency now has entirely new definitions, taking factors like environmental cost into consideration over dialogue-tree depth. Proactivity is achievable, but still through narrowing user autonomy. Personalization is also finally possible, but really puts privacy at the forefront in a way that pre-LLM CAs never had to consider because they simply never had enough personal information or personality to threaten it. The breakdowns we now observe in post-LLM deployments include hallucinated or ungrounded expectations and ambiguous check-in cadences. We have gone from frustration at not being able to "speak CA" to questions about whether the user has to do anything at all. So what does success mean if the system is designed to require nothing of the user? Our pre-LLM metrics: interaction counts, task completion and task categorization were built for the paradigm in which the user had to act first. They break down in the 2.0 paradigm where the most successful interaction may be the one the user never had to initiate at all, or where a long multi-turn conversation brings a smile to a lonely face. Most importantly, our metrics across both eras have overwhelmingly served research convenience and researcher definitions of success, rather than user-defined metrics. Our participants have never described their interactions in terms of counts or completion but rather in terms of relief, companionship, trust and independence. To this end, our single biggest recommendation and learning to date is to center qualitative feedback above quantitative feedback, even if that is harder for the researcher. As CAs become capable of open-ended, explanatory, proactive interaction that older adults seem to have wanted all along, our obligation as researchers is to let our metrics catch up to not only as “how often” or “how successfully” a system was used, but what it actually meant to the person using it. 

\bibliographystyle{ACM-Reference-Format}
\bibliography{GROUP-27}

\end{document}